# Near-field radiative heat transfer between dissimilar materials mediated by coupled surface phonon- and plasmon-polaritons

Lei Tang[†], John DeSutter[‡], and Mathieu Francoeur[‡,*]

[†]Department of Mechanical Engineering, University of California, Berkeley, Berkeley, CA 94720, USA

[‡]Radiative Energy Transfer Lab, Department of Mechanical Engineering, University of Utah, Salt Lake City, UT 84112, USA

**ABSTRACT:** Near-field radiative heat transfer (NFRHT) between dissimilar materials supporting surface polaritons in the infrared is of critical importance for applications such as photonic thermal rectification and near-field thermophotovoltaics. Here, we measure NFRHT between millimeter-size surfaces made of 6*H*-SiC and doped Si, respectively supporting surface phonon-polaritons (SPhPs) and surface plasmon-polaritons (SPPs) in the infrared, separated by a 150-nm-thick vacuum gap spacing maintained via $SiO_2$ nanopillars. For purpose of comparison, measurements are also performed between two doped Si surfaces. The measured radiative flux is in good agreement with theoretical predictions based on fluctuational electrodynamics. A flux enhancement beyond the blackbody limit of ~ 8.2 is obtained for the SiC-Si sample, which is smaller than the enhancement for the Si-Si sample (~ 12.5) owing to the spectral mismatch of the SiC and Si light lines, and SPhP and SPP resonances. However, due to lower losses in SiC than Si and weaker SPhP-SPP coupling than SPP coupling, the near-field enhancement for the SiC-Si sample exhibits a more pronounced monochromatic behavior with a resonant flux that is ~ 5 times larger than the resonant flux for the Si-Si sample. This work demonstrates that it is possible to modulate NFRHT via surface polariton coupling, and will accelerate the development of energy conversion and thermal management devices capitalizing on the near-field effects of thermal radiation between dissimilar materials.

**KEYWORDS:** *near-field radiative heat transfer, radiative flux measurement, dissimilar materials, coupled surface phonon- and plasmon-polaritons, silicon carbide, doped silicon*

Near-field radiative heat transfer (NFRHT), arising when heat sources are separated by subwavelength gap spacings, can significantly exceed Planck's blackbody limit owing to tunneling of evanescent waves.[1-6] These evanescent waves include broadband frustrated modes and narrowband surface polaritons that can lead to quasi-monochromatic radiative flux.[7] Many potential applications of NFRHT, such as near-field thermophotovoltaics,[8-12] photonic thermal rectification[13,14] and flux modulation,[15] capitalize on the coupling of surface polaritons between dissimilar materials. Measurements of NFRHT between dissimilar materials in the microsize sphere-surface ($SiO_2$-Si,[16] $SiO_2$-Au[17]) and microsize mesa-surface ($SiO_2$-Au,[18] Si-$VO_2$[14]) configurations have been performed. However, from an application standpoint, the ability to demonstrate NFRHT between macroscale surfaces (~ $mm^2$) is critical, since the heat transfer rate is not only proportional to the near-field enhancement of the radiative flux, but also to the size of the surfaces. NFRHT measurements between macroscale surfaces made of the same materials, namely Si,[19-22] $SiO_2$,[23-27] $Al_2O_3$,[28] Al,[29] graphene,[30] graphene-covered $SiO_2$[31], and metallo-dielectric multilayers[32] have been reported. Only Ito et al.[33] measured NFRHT between macroscale surfaces made of dissimilar materials ($VO_2$-$SiO_2$), but the vacuum gap spacing of 370 nm was too large for NFRHT to be dominated by coupled surface polaritons.

In this paper, we address a critical knowledge gap by measuring NFRHT between macroscale surfaces made of dissimilar materials. Specifically, $5 \times 5$ $mm^2$ surfaces of 6*H*-SiC and doped Si are selected as they respectively support surface phonon-polaritons (SPhPs) and surface plasmon-polaritons (SPPs) near a wavelength of 10 μm that can be thermally excited at room temperature. The surfaces are separated by a 150-nm-thick vacuum gap spacing via $SiO_2$ nanopillars. By comparing experimental results in the SiC-Si configuration against those obtained between two Si surfaces, it is shown that NFRHT mediated by coupled SPhPs and SPPs leads to a lower radiative flux (enhancement of ~ 8.2 beyond the blackbody limit) than SPP mediated NFRHT (enhancement of ~ 12.5 beyond the blackbody limit) owing to the spectral mismatch of SPhP and SPP resonances and light lines. However, coupled SPhP and SPP mediated NFRHT shows a more pronounced monochromatic behavior with a larger resonant radiative flux (~ 5 times larger than for Si-Si) due to a weaker surface polariton coupling in the vacuum gap spacing and lower losses in SiC.

**EXPERIMENTAL PROCEDURE**

The samples used for measuring NFRHT consist of a high-temperature emitter and a low-temperature receiver, both characterized by surface areas of 5 × 5 $mm^2$, separated by a 150-nm-thick gap spacing *d* maintained via five rigid $SiO_2$ nanopillars. The 3-µm-diameter $SiO_2$ nanopillars are manufactured onto the 525-µm-thick receiver made of Si with boron doping of ~ $4.6 \times 10^{19}$ $cm^{-3}$ (see Figure 1(a)). The nanopillars cover only $1.4 \times 10^{-4}$ % of the receiver surface. The emitter is made of a 315-µm-thick 6*H*-SiC substrate. In order to compare coupled SPhP and SPP against SPP mediated NFRHT, a sample with an emitter of Si having the same doping level as the receiver is manufactured and tested. The Si (Silicon Valley Microelectronics) and SiC (MTI, SC6HZ050503-033s1) substrates are respectively characterized by surface roughness less than 0.2 and 1 nm, as provided by the manufacturers and as measured via a Zygo NewView 5000 optical profilometer. The emitter is manually deposited onto the $SiO_2$ nanopillars and aligned with the receiver without applying any external pressure. Fabrication of the NFRHT samples is discussed in Supporting Information (SI) Section 1. The vacuum gap spacing of the samples have been characterized ex situ by measuring the nanopillar heights using a Tencor P-20H profilometer and by performing a structural analysis for determining the sample bow and nanopillar deflection (see SI Section 2). For the SiC-Si and Si-Si samples, the vacuum gap spacings, *d*, are respectively estimated to be $150^{+68}_{-24}$ nm and $150^{+56}_{-24}$ nm.

Figure 1(b) shows a schematic of the setup used to perform NFRHT measurements. The temperature difference between the emitter and receiver is maintained via two thermoelectric (TE) modules. The top TE (Custom Thermoelectric, 00701-9B30-22RU4) acts as a heater by providing a heat rate $Q$ to the sample, which is the sum of the power supplied by the TE, $P_h$, and the power into the TE, $Q_{in}$. The experiments are performed under vacuum conditions, such that the contribution of $Q_{in}$, due to thermal emission by the walls of the vacuum chamber, is negligible compared to $P_h$. The bottom TE (TETechnology, VT-31-1.0-1.3) is used to maintain the receiver at a temperature of ~ 300 K by directing heat towards a large Cu heat sink. The total heat rate flowing in the sample, $Q$, includes NFRHT between the emitter and receiver across the vacuum gap spacing, $Q_{rad}$, and conduction heat transfer through the $SiO_2$ nanopillars, $Q_{cond}$, characterized by a thermal conductivity of 1.3 $Wm^{-1}K^{-1}$ at room temperature.[34] The heat rate, $Q$, is measured via a heat flux meter (Fluxteq, PHFS-JD10) located between the receiver and the TE cooler. To ensure a uniform heat flux across the surfaces, the heat flux meter is surrounded by two, 500-µm-thick Cu heat spreaders. The high and low temperatures, $T_h$ and $T_l$, are measured via two thermistors (Selco, LSMC700A010KD002) that are embedded inside two 500-µm-thick Cu heat spreaders adjacent to the emitter and receiver. Note that all layers in the experimental setup have surface areas of $5 \times 5$ mm$^2$, except the heat flux meter (and the surrounding Cu heat spreaders) and the TE cooler that are characterized by surface areas of $10 \times 10$ mm$^2$ and $14 \times 14$ mm$^2$, respectively. Although not shown in the schematic, thermal grease (Arctic Silver Ceramique 2) is applied to all interfaces in order to minimize the thermal contact resistances. In addition, a force is applied on the TE heater via a calibrated 10-g-mass to further reduce the thermal contact resistances. The temperatures adjacent to the vacuum gap spacing, $T_e$ and $T_r$, needed for the theoretical predictions of NFRHT and conduction through the nanopillars are retrieved via the temperatures measured with the two thermistors, $T_h$ and $T_l$, the measured thermal resistance of the grease at the emitter-Cu and receiver-Cu interfaces, $R_g$, and the thermal resistances of the emitter, $R_e$, and receiver, $R_r$ (see the equivalent thermal circuit in Figure 1(c)). The measured thermal resistance of the grease is less than ~ 6.2 K/W,[35] while the thermal resistances by conduction of the SiC (thermal conductivity of 490 $Wm^{-1}K^{-1}$)[36] and Si (thermal conductivity of 150 $Wm^{-1}K^{-1}$)[36] substrates are respectively ~ 0.026 K/W and ~ 0.14 K/W. The thermal resistances of the emitter and receiver are thus negligibly small compared to the thermal resistances of the grease and of the vacuum gap spacing that includes NFRHT between the emitter and the receiver and conduction through the $SiO_2$ nanopillars. For a 150-nm-thick gap spacing, and emitter and receiver temperatures of 370 K and 300 K, the thermal resistance in the gap spacing is ~ 571 K/W. As such, the temperatures across the emitter and receiver are assumed to be uniform, as $T_e$ and $T_r$ are retrieved from $T_h$ and $T_l$ using solely the thermal resistance of the grease. The heat flux meter has been calibrated by measuring the thermal resistance of 1.1-mm-thick borosilicate glass having a known thermal conductivity of 0.94 $Wm^{-1}K^{-1}$.[21] All experiments have been conducted in a vacuum chamber with a pressure of ~ $10^{-4}$ Pa under a class 1000 clean room tent.

## RESULTS AND DISCUSSION

The radiative flux $q_{rad}$ plotted in Figure 2 as a function of the temperature difference, $\Delta T = T_e - T_r$, is retrieved by dividing the heat rate due to NFRHT $Q_{rad}$, obtained by subtracting conduction through the $SiO_2$ nanopillars $Q_{cond}$ from the measured heat rate $Q$, by the sample surface area. The results are compared against fluctuational electrodynamics predictions and the radiative flux between two blackbodies (see Methods for radiation and conduction calculations). The colored bands of theoretical predictions are calculated based on the nanopillar height, sample bow and Si doping level (see SI Section 3 for uncertainty analysis). The measured radiative flux is in good agreement with fluctuational electrodynamics predictions for both the SiC-Si and Si-Si samples. The unprocessed measured heat rate, $Q$, which include NFRHT across the vacuum gap spacing and conduction through the $SiO_2$ nanopillars as a function of the temperature difference, $\Delta T$, is provided in SI Section 3 (Figure S6). For both samples, the heat rate $Q$ is largely dominated by NFRHT. For a temperature difference of 70 ± 3 K, the relative contribution of conduction through the $SiO_2$ nanopillars, $Q_{cond}$, to the overall measured heat rate, $Q$, takes maximum and minimum values of 17.6% and 12.7% for the SiC-Si sample, and 12.4% and 8.9% for the Si-Si sample.

The measured radiative flux for the SiC-Si and Si-Si samples respectively exceeds the blackbody limit by factors of ~ 8.2 and ~ 12.5 for a temperature difference of 70 ± 3 K. The physics underlying the near-field enhancement is explained by analyzing the radiative flux in transverse magnetic (TM) and transverse electric (TE) polarizations per unit angular frequency, $\omega$, and per unit parallel wavevector, $k_\rho$, for a temperature difference of 70 K (see Figure 3). The vacuum light line ($k_\rho = k_0$), the material light line ($k_\rho = \sqrt{|\varepsilon_j|}k_0$, where $\varepsilon_j$ is the dielectric function of medium $j$), and the surface polariton dispersion relation are shown in Figure 3 (see Methods for calculation of the material light line and surface polariton dispersion relation). Note that surface polariton dispersion relations are shown only in TM polarization, since these modes can only be excited in that polarization state for non-magnetic materials.[37]

For the Si-Si sample, the contribution of propagating modes ($k_\rho < k_0$), propagating in both Si and vacuum, is modest and accounts for ~ 4.8% of the total radiative flux. Frustrated modes ($k_0 < k_\rho < \sqrt{|\varepsilon_{Si}|}\, k_0$), which are propagating in Si and evanescent in vacuum, contribute to the total radiative flux in a larger proportion (~ 35.2%). The largest contribution comes from SPPs ($k_\rho > \sqrt{|\varepsilon_{Si}|}\, k_0$), that are evanescent in both Si and vacuum, accounting for ~ 60% of the total radiative flux. The SPPs supported by the emitter and receiver couple in the vacuum gap spacing, and split the dispersion relation into a high-frequency antisymmetric mode, $\omega^+$, and a low-frequency symmetric mode, $\omega^-$. For large $k_\rho$ values, the antisymmetric and symmetric modes converge to the resonant frequency of a Si-vacuum interface, $\omega_{SPP}$, of 1.765 × $10^{14}$ rad/s (wavelength of 10.7 μm). The radiative flux around SPP dispersion relation spreads out over a large spectral band owing to high losses in Si (the imaginary part of the dielectric function of Si takes values of ~ 19.6 and ~ 3.62 for frequencies of 1 × $10^{14}$ rad/s and 2 × $10^{14}$ rad/s, respectively). This results in a spectral radiative flux (i.e., the radiative flux integrated over all wavevectors $k_\rho$) that is maximum near the resonant frequency of a Si-vacuum interface and that spreads out over frequencies of ~ 1 × $10^{14}$ rad/s to 2 × $10^{14}$ rad/s, corresponding to wavelengths of ~ 9.42 μm to 18.8 μm (see Figure 4(a)). Note that the maximum in Figure 4(a) (~ 1.624 × $10^{14}$ rad/s) is not exactly equal to $\omega_{SPP}$ as losses are neglected when estimating the resonant frequency of a Si-vacuum interface.

The physics of NFRHT between dissimilar materials is more complex due to the emitter and receiver having different material light lines, and due to the spectral mismatch of SPhP and SPP resonances at the SiC-vacuum and Si-vacuum interfaces. While the light line in Si is always larger than the vacuum light line, the light line in SiC is smaller than the vacuum light line in a portion of the Reststrahlen band delimited by the transverse (1.476 × $10^{14}$ rad/s) and longitudinal (1.804 × $10^{14}$ rad/s) optical phonon frequencies. In that case, propagating modes cannot contribute to NFRHT when the parallel wavevector $k_\rho$ is smaller than $k_0$ but larger than $\sqrt{|\varepsilon_{\mathrm{SiC}}|}k_0$ as these modes cannot propagate in SiC. In the rest of the Reststrahlen band, propagating modes do not contribute to the radiative flux due to the highly metallic behavior of SiC (real part of the dielectric function is negative). This can be seen in Figure 3 for the TE polarization state and in Figure 4(b) for propagating modes. Outside the Reststrahlen band, propagating modes contribute to NFRHT in a modest manner as they account for ~ 10.5% of the total radiative flux.

As for propagating modes, frustrated modes are not supported in the spectral band where the parallel wavevector simultaneously satisfies $k_\rho < k_0$ and $k_\rho > \sqrt{|\varepsilon_{\text{SiC}}|}k_0$. Outside that spectral band, frustrated modes are limited to $k_\rho = \min(\sqrt{|\varepsilon_{\text{SiC}}|}k_0, \sqrt{|\varepsilon_{\text{Si}}|}k_0)$. Note that between frequencies of ~ $1.2 \times 10^{14}$ to $1.4 \times 10^{14}$ rad/s, the radiative flux due to frustrated modes is non-zero for $k_\rho$ values slightly exceeding $\min(\sqrt{|\varepsilon_{\text{SiC}}|}k_0, \sqrt{|\varepsilon_{\text{Si}}|}k_0)$; this is better seen in TE polarization in Figure 3 where surface polaritons do not contribute to NFRHT. This is explained by the large losses in Si where the imaginary part of its dielectric function takes values of ~ 13.1 and ~ 9.05 for frequencies of $1.2 \times 10^{14}$ rad/s and $1.4 \times 10^{14}$ rad/s, respectively. For the SiC-Si sample, the contribution of frustrated modes to NFRHT accounts for ~ 54.2% of the total radiative heat flux. The radiative flux from frustrated modes for the Si-Si sample (~ 2776 W/m$^2$) is ~ 1.3 times larger than for the SiC-Si sample (~ 2192 W/m$^2$) when the temperature difference is 70 K. This can be explained by the spectral mismatch of the SiC and Si light lines. As shown in Figure 3, the SiC light line takes smaller $k_\rho$ values than the Si light lines for most frequencies, which limits the contribution of frustrated modes to the radiative flux.

SiC supports SPhPs in the Reststrahlen band, and the resonant frequency of a SiC-vacuum interface, $\omega_{SPhP}$, is ~ $1.765 \times 10^{14}$ rad/s (wavelength of 10.7 μm). Here, $\omega_{SPhP}$ is the same as the calculated SPP resonant frequency of a Si-vacuum interface, but is slightly different than the flux resonance of ~ $1.624 \times 10^{14}$ rad/s shown in Figure 4(a). SPhPs and SPPs couple in the vacuum gap spacing and split into antisymmetric ($\omega^+$) and symmetric ($\omega^-$) modes. For large $k_\rho$ values, the antisymmetric and symmetric modes converge to a single frequency of ~ $1.765 \times 10^{14}$ rad/s owing to coupled SPhPs and SPPs, and this prediction agrees well with the resonance of the spectral radiative flux shown in Figure 4(b) (~ $1.764 \times 10^{14}$ rad/s). Indeed, SiC has small losses near SPhP resonant frequency (imaginary part of the dielectric function smaller than 1), such that it is safe to model SiC as lossless when calculating the dispersion relation. Coupled SPhPs and SPPs only contribute to ~ 35.3% of the total radiative flux. The contribution of surface polaritons to the total radiative flux is smaller for the SiC-Si sample than for the Si-Si sample for two reasons. First, owing to weaker coupling, the antisymmetric and symmetric modes for the SiC-Si sample converge into a single resonance for smaller $k_\rho$ values than for the Si-Si sample, thus resulting in surface polariton mediated NFRHT occurring within a narrower spectral band. Second, compared to Si, SiC is characterized by low losses near SPhP resonant frequency. As such, the radiative flux spreads out over a much narrower spectral band around SPhP-SPP dispersion relation than around SPP dispersion relation for the Si-Si sample. This can be clearly seen by comparing the results for the SiC-Si and Si-Si samples in TM polarization shown in Figure 3, and by inspecting Figure 4(b) where coupled SPhPs and SPPs result in a sharp, narrowband enhancement of the spectral radiative flux. Therefore, the radiative flux is smaller for coupled SPhPs and SPPs than for coupled SPPs by a factor of ~ 3.3 (~ 4718 W/m$^2$ for Si-Si and ~ 1427 W/m$^2$ for SiC-Si). However, the resonance of the radiative flux is significantly larger for the SiC-Si sample than for the Si-Si sample (~ 5 times larger).

## CONCLUSIONS

In summary, this work experimentally investigated NFRHT between dissimilar materials made of $5 \times 5$ mm$^2$ 6*H*-SiC and doped Si surfaces respectively supporting SPhPs and SPPs in the infrared. The surfaces were separated by a 150-nm-thick vacuum gap spacing maintained via $SiO_2$ nanopillars. Experimental results for NFRHT between two doped Si surfaces have also been reported for purpose of comparison. The experimental results were in good agreement with fluctuational electrodynamics predictions, and near-field flux enhancements of ~ 8.2 and ~ 12.5 beyond the blackbody limit have been respectively measured for the SiC-Si and Si-Si samples. Despite a smaller enhancement, SPhP and SPP mediated NFRHT with the SiC-Si sample results in a more pronounced monochromatic behavior with a flux resonance that is ~ 5 times larger the flux resonance for the Si-Si sample capitalizing on coupled SPPs. This monochromatic behavior resulting from surface polariton coupling between dissimilar materials demonstrates the ability of controlling NFRHT, which is critical for many applications such as photonic thermal rectification and near-field thermophotovoltaics.

## METHODS

**Radiation and conduction calculations.** In the radiation calculations, the emitter and receiver are modeled as semi-infinite layers. Using fluctuational electrodynamics,[38] the contributions of propagating, frustrated, and surface polariton modes to the total (i.e., spectrally integrated) radiative flux between the emitter ($e$) and the receiver ($r$) are respectively calculated as follows:

$$q_{rad}^{prop} = \frac{1}{4\pi^2}\int_0^\infty [\Theta(\omega,T_e)-\Theta(\omega,T_r)]d\omega \left[\int_0^{k_0} k_\rho\, dk_\rho \sum_{\gamma=TE,TM} \frac{\left(1-\left|r_{0e}^\gamma\right|^2\right)\left(1-\left|r_{0r}^\gamma\right|^2\right)}{\left|1-r_{0e}^\gamma r_{0r}^\gamma e^{2i\mathrm{Re}(k_{z0})d}\right|^2}\right] \tag{1}$$

$$q_{rad}^{frus} = \frac{1}{\pi^2}\int_0^\infty [\Theta(\omega,T_e)-\Theta(\omega,T_r)]d\omega \left[\int_{k_0}^{\sqrt{|\varepsilon_{\mathrm{Si/SiC}}|}k_0} k_\rho dk_\rho e^{-2k''_{z0}d} \sum_{\gamma=TE,TM} \frac{\mathrm{Im}\left(r_{0e}^\gamma\right)\mathrm{Im}\left(r_{0r}^\gamma\right)}{\left|1-r_{0e}^\gamma r_{0r}^\gamma e^{-2\mathrm{Im}(k_{z0})d}\right|^2}\right] \tag{2}$$

$$q_{rad}^{SP} = \frac{1}{\pi^2}\int_0^\infty [\Theta(\omega,T_e)-\Theta(\omega,T_r)]d\omega \left[\int_{\sqrt{|\varepsilon_{\mathrm{Si/SiC}}|}k_0}^{\infty} k_\rho\, dk_\rho e^{-2k''_{z0}d} \sum_{\gamma=TE,TM} \frac{\mathrm{Im}\left(r_{0e}^\gamma\right)\mathrm{Im}\left(r_{0r}^\gamma\right)}{\left|1-r_{0e}^\gamma r_{0r}^\gamma e^{-2\mathrm{Im}(k_{z0})d}\right|^2}\right] \tag{3}$$

where the subscript 0 refers to vacuum, $\gamma$ denotes the polarization state (TE or TM), and $k_{z0}$ is the component of the vacuum wavevector perpendicular to the surfaces. The mean energy of an electromagnetic state, $\Theta(\omega,T)$, is given by:

$$\Theta(\omega,T)=\frac{\hbar\omega}{\exp(\hbar\omega/k_BT)-1} \tag{4}$$

where $\hbar$ is the reduced Planck constant (= 1.055 × $10^{-34}$ Js) and $k_B$ is the Boltzmann constant (= 1.381 × $10^{-23}$ J/K). In Eqs. (1)-(3), $r_{0j}^{\gamma}$ is the Fresnel reflection coefficient at the vacuum-material interface. The Fresnel reflection coefficients in TE and TM polarizations are respectively calculated following:

$$r_{0j}^{TE}=\frac{k_{z0}-k_{zj}}{k_{z0}+k_{zj}} \tag{5}$$

$$r_{0j}^{TM}=\frac{\varepsilon_j k_{z0}-k_{zj}}{\varepsilon_j k_{z0}+k_{zj}} \tag{6}$$

where $\varepsilon_j$ is the dielectric function of medium *j*. The temperature-dependent dielectric function of 6*H*-SiC is expressed as[39,40]:

$$\varepsilon_{\mathrm{SiC}}(\omega,T)=\varepsilon_{\mathrm{SiC},\infty}\left(\frac{\omega^2-\omega_{LO}^2+i\Gamma\omega}{\omega^2-\omega_{TO}^2+i\Gamma\omega}\right) \tag{7}$$

where the damping factor Γ takes a constant value of 1.036 × $10^{12}$ rad/s. The temperature-dependent high-frequency dielectric constant $\varepsilon_{\mathrm{SiC},\infty}$, longitudinal optical phonon frequency $\omega_{LO}$, and the transverse optical phonon frequencies $\omega_{TO}$ are respectively given by:

$$\varepsilon_{\mathrm{SiC},\infty}=6.7\exp\left[2.5\times10^{-5}\left(T-300\right)\right] \tag{8}$$

$$\omega_{LO}=1.808\times10^{14}-5.839\times10^{9}\left(T-300\right)\ \text{rad/s} \tag{9}$$

$$\omega_{TO}=1.48\times10^{14}-5.651\times10^{9}\left(T-300\right)\ \text{rad/s} \tag{10}$$

Note that 6*H*-SiC is described by anisotropic optical properties. However, since the temperature-dependent dielectric functions in the directions along the extraordinary and ordinary axes are very similar, 6*H*-SiC is assumed to be isotropic and only the dielectric function along the extraordinary axis has been used for generating the results in this work.

The temperature-dependent dielectric function of doped Si is described by a Drude model[41,42]:

$$\varepsilon_{\mathrm{Si}}(\omega,T)=\varepsilon_{\mathrm{Si},\infty}-\frac{\omega_p^2}{\omega(\omega+i\gamma)} \tag{11}$$

where $\varepsilon_{\mathrm{Si},\infty}$ = 11.7 is the high-frequency dielectric constant. The plasma frequency $\omega_p$ and scattering rate $\gamma$ are respectively calculated as follows:

$$\omega_p=\sqrt{\frac{N_h e^2}{m^* \varepsilon_0}} \tag{12}$$

$$\gamma=\frac{e}{m^* \mu} \tag{13}$$

where $N_h$ is the temperature-dependent hole concentration, $e$ is the electron charge, $m^*$ is the hole effective mass, $\varepsilon_0$ is the vacuum permittivity, and $\mu$ is the temperature-dependent mobility. The temperature-dependent expressions for the hole concentration and mobility can be found in Refs. 41 and 42.

When both the emitter and receiver are modeled as blackbodies, the total radiative flux is calculated via:

$$q_b=\pi\int_0^\infty [I_{b,\omega}(T_e)-I_{b,\omega}(T_r)]d\omega \tag{14}$$

where the spectral blackbody intensity is defined as:

$$I_{b,\omega}(T)=\frac{\hbar\omega^3}{4\pi^3 c_0^2[\exp(\hbar\omega / k_B T)-1]} \tag{15}$$

Steady-state, one-dimensional conduction heat transfer through the nanopillars separating the emitter and receiver is assumed. This is justified by the fact that the nanopillar temperature parallel to the emitter and receiver surfaces is nearly uniform. The heat rate by conduction is estimated from Fourier's law where the temperature in the nanopillars varies linearly in the direction normal to the emitter and receiver surfaces:

$$Q_{cond} = \kappa NA\frac{T_e - T_r}{d} \tag{16}$$

where $\kappa$ is the thermal conductivity of the $SiO_2$ nanopillars, $N$ is the number of nanopillars, and $A$ is the contact area between a nanopillar and the emitter/receiver. The contact area $A$ is assumed to be equal to the nanopillar cross-sectional area. A temperature-independent thermal conductivity of 1.3 $Wm^{-1}K^{-1}$ is used in the calculations for $SiO_2$.[34]

The total heat rate flowing through a sample $Q$, due to NFRHT across the vacuum gap spacing and conduction through the $SiO_2$ nanopillars, is calculated by multiplying the sum of Eqs. (1), (2) and (3) by the sample surface area (5 × 5 $mm^2$), and by adding Eq. (16) to the result.

**Material light line.** Electromagnetic waves propagating in medium $j$ described by a dielectric function $\varepsilon_j$ must satisfy the following dispersion relation:

$$k_\rho^2 + k_{zj}^2 = \varepsilon_j k_0^2 \tag{17}$$

From Eq. (17), it is concluded that the maximum parallel wavevector beyond which waves cannot propagate in medium $j$ is $k_\rho^2 = \varepsilon_j k_0^2$, which corresponds to the material light line. Care must be taken when evaluating the material light line, as the left-hand side of the last equality is a pure real number while the right-hand side is complex. As such, the material light can be approximated by taking the magnitude of the right-hand side of the equality to ensure that the parallel wavevector is a pure real number[43]:

$$k_\rho \approx \sqrt{|\varepsilon_j|}k_0 \tag{18}$$

In the limit that losses are negligible in medium $j$, Eq. (18) reduces to $k_\rho \approx \mathrm{Re}\left(\sqrt{\varepsilon_j}\right)k_0$, which is the expression that is typically used for calculating the material light line[7].

It is important to emphasize that it is impossible to perfectly define the material light line when there are losses in medium $j$. In this work, Eq. (18) is used for calculating all material light lines, as it provides better results than $k_\rho \approx \mathrm{Re}\left(\sqrt{\varepsilon_j}\right)k_0$ when losses are large.

**Surface polariton dispersion relation.** The antisymmetric, $\omega^+$, and symmetric, $\omega^-$, modes plotted in Figure 3 are obtained by numerically solving $1 - r_{01}^{\gamma} r_{02}^{\gamma} e^{-2k''_{z0}d} = 0$ in TM polarization and by neglecting losses in the dielectric function of SiC and Si[44]. Since the dielectric functions of both SiC and Si are temperature dependent, these modes are calculated for an emitter temperature of 370 K and a receiver temperature of 300 K. In the electrostatic limit where $k_\rho >> k_0$, the antisymmetric and symmetric modes converge to a single resonant frequency. The resonant frequencies of single SiC-vacuum[45] and Si-vacuum[37] interfaces are respectively given by:

$$\omega_{SPhP} \approx \sqrt{\frac{\varepsilon_{\mathrm{SiC},\infty}\omega_{LO}^2 + \omega_{TO}^2}{\varepsilon_{\mathrm{SiC},\infty} + 1}} \tag{19}$$

$$\omega_{SPP} \approx \frac{\omega_p}{\sqrt{\varepsilon_{\mathrm{Si},\infty} + 1}} \tag{20}$$

## ASSOCIATED CONTENT

The Supporting Information is available free of charges at [URL to be included by the publisher].

Steps detailing sample fabrication, gap spacing estimation, and uncertainty analysis.

## AUTHOR INFORMATION

### Corresponding Authors

*E-mail: mfrancoeur@mech.utah.edu

**ORCID**

Mathieu Francoeur: 0000-0003-4989-4861

**Author Contributions**

This work was conceived by L.T., J.D. and M.F. Design, fabrication and testing of the samples were performed by L.T. and J.D. under the supervision of M.F. Numerical simulations were performed by L.T. under the supervision of M.F. The manuscript was written by L.T. and M.F with comments provided by J.D.

**Notes**

The authors declare no competing financial interest.

**ACKNOWLEDGEMENTS**

The authors acknowledge financial support from the National Science Foundation (grant no. CBET-1253577). This work was performed in part at the Utah Nanofab sponsored by the College of Engineering, Office of the Vice President for Research and the Utah Science Technology and Research (USTAR) initiative of the State of Utah. The authors appreciate the support of the staff and facilities that made this work possible. This work also made use of University of Utah shared facilities of the Micron Technology Foundation Inc. Microscopy Suite sponsored by the College of Engineering, Health Sciences Center, Office of Vice President for Research and the Utah Science Technology and Research (USTAR) initiative of the State of Utah.

## FIGURES

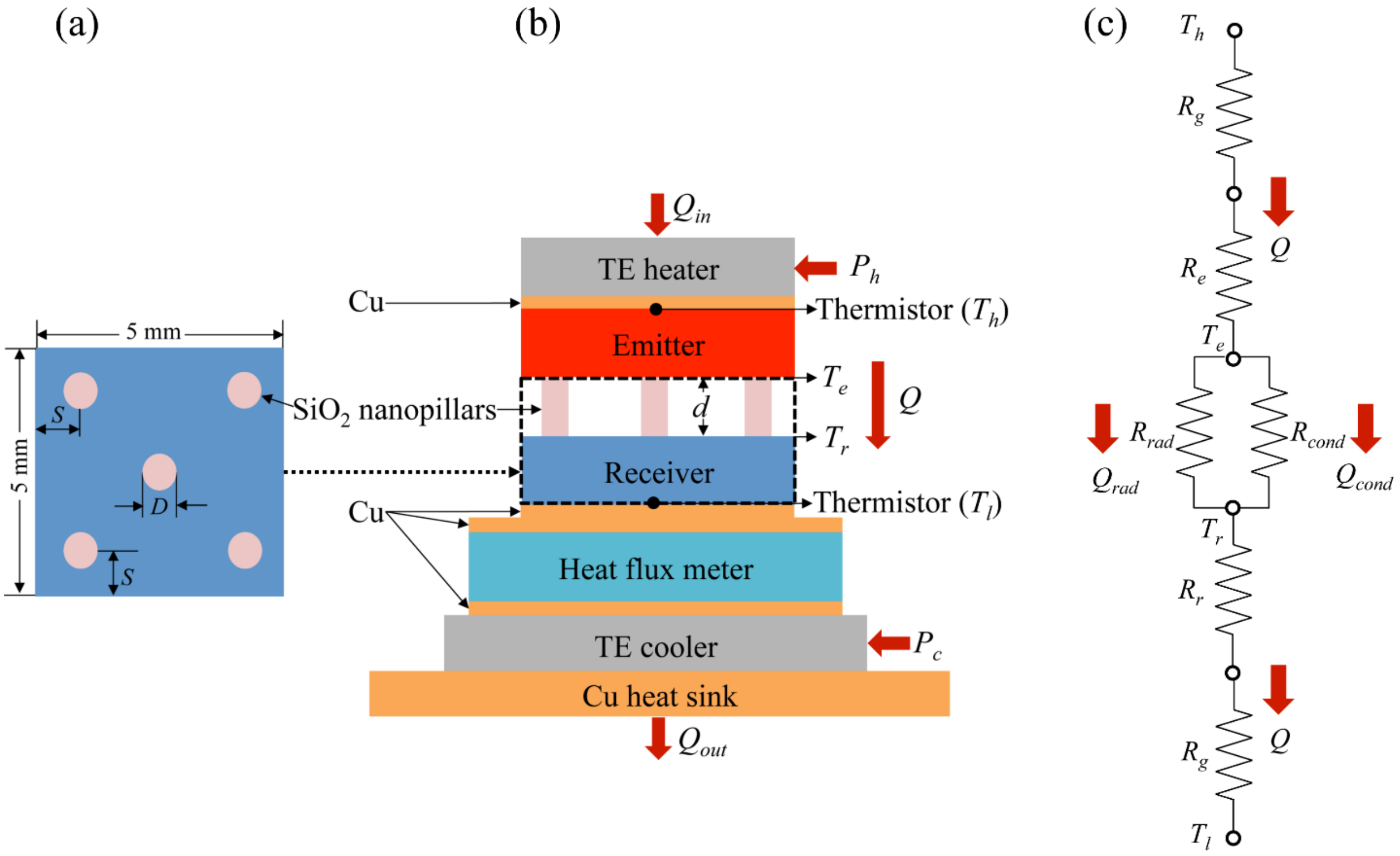

**Figure 1.** NFRHT sample and measurement setup. (a) $SiO_2$ nanopillars manufactured onto the Si receiver. The nanopillars are characterized by diameters $D$ of 3 μm. One nanopillar is located in the center of the receiver, while the four other nanopillars are located at distances $S$, measured with respect to the edges of the receiver, of 1 mm (SiC-Si sample) and 1.5 mm (Si-Si sample). (b) NFRHT measurement setup. The SiC-Si and Si-Si samples are characterized by $5 \times 5$ mm$^2$ surfaces and are separated by a vacuum gap spacing of 150 nm via five $SiO_2$ nanopillars. The heat rate through the device, $Q$, is measured via a heat flux meter and includes NFRHT between the emitter and receiver across the vacuum gap spacing and conduction through the $SiO_2$ nanopillars. The emitter and receiver temperatures, $T_e$ and $T_r$, are retrieved from the measured temperatures $T_h$ and $T_l$ and the thermal resistance of the grease. (c) Equivalent thermal circuit of the setup. $R_g$, $R_e$ and $R_r$ are the thermal resistances of the grease, the emitter, and the receiver, while $R_{rad}$ and $R_{cond}$ are the thermal resistances due to NFRHT in the vacuum gap spacing and conduction through the $SiO_2$ nanopillars.

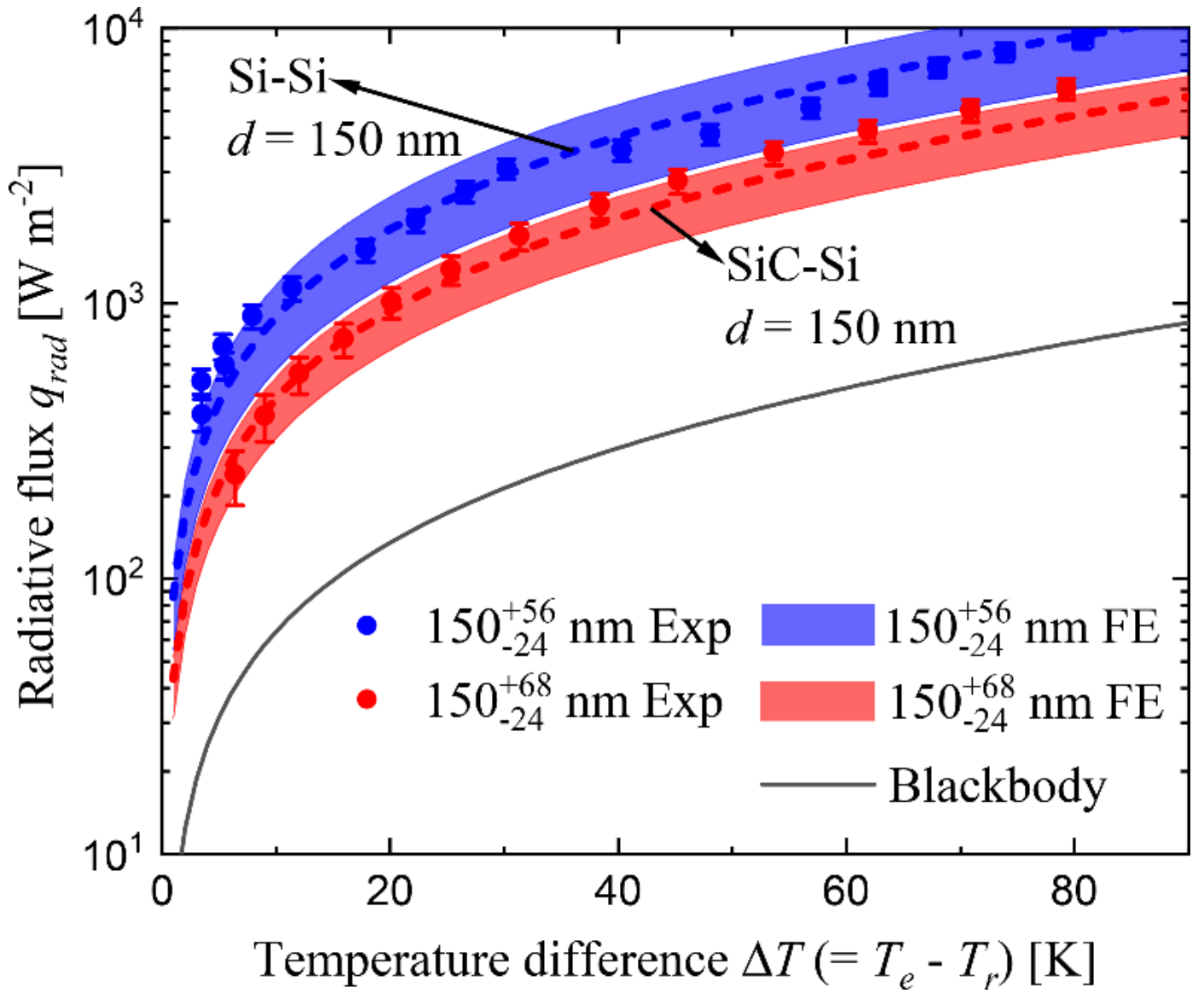

**Figure 2.** Radiative heat flux, $q_{rad}$, as a function of the temperature difference, $\Delta T = T_e - T_r$, for the SiC-Si and Si-Si samples. Experimental results (Exp) are shown by symbols, where conduction heat transfer through the $SiO_2$ nanopillars has been subtracted, and are compared against fluctuational electrodynamics (FE) and blackbody predictions. The colored bands of FE predictions are calculated based on the nanopillar height, sample bow and Si doping level. In all experiments, the temperature of the receiver, $T_r$, is kept at ~ 300 K.

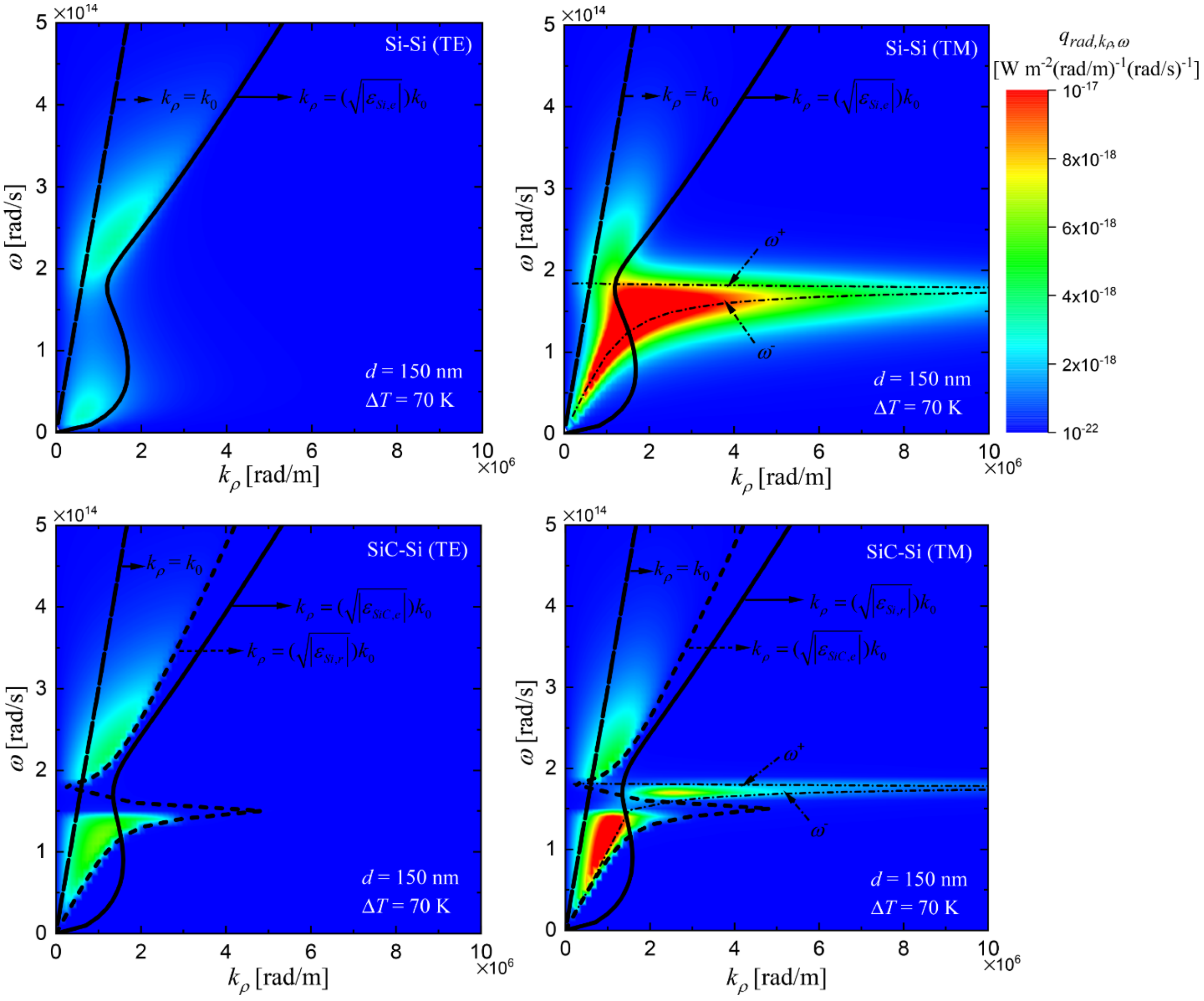

**Figure 3.** Radiative heat flux, $q_{rad,k_\rho,\omega}$, per unit angular frequency, $\omega$, and per unit parallel wavevector, $k_\rho$, for a vacuum gap spacing of 150 nm and a temperature difference of 70 K ($T_e$ = 370 K, $T_r$ = 300 K). The radiative flux is shown for the SiC-Si and Si-Si samples in TE and TM polarization states. The vacuum light line is identified by a dashed line, while the material light lines are plotted as short-dashed (SiC) and solid (Si) lines. For the Si-Si sample, only the light line in the emitter at 370 K is shown since the receiver light line at 300 K is essentially the same. The antisymmetric, $\omega^+$, and symmetric, $\omega^-$, modes calculated from surface polariton dispersion relations are also plotted. For purpose of comparison, all panels share the same color scale.

(a)

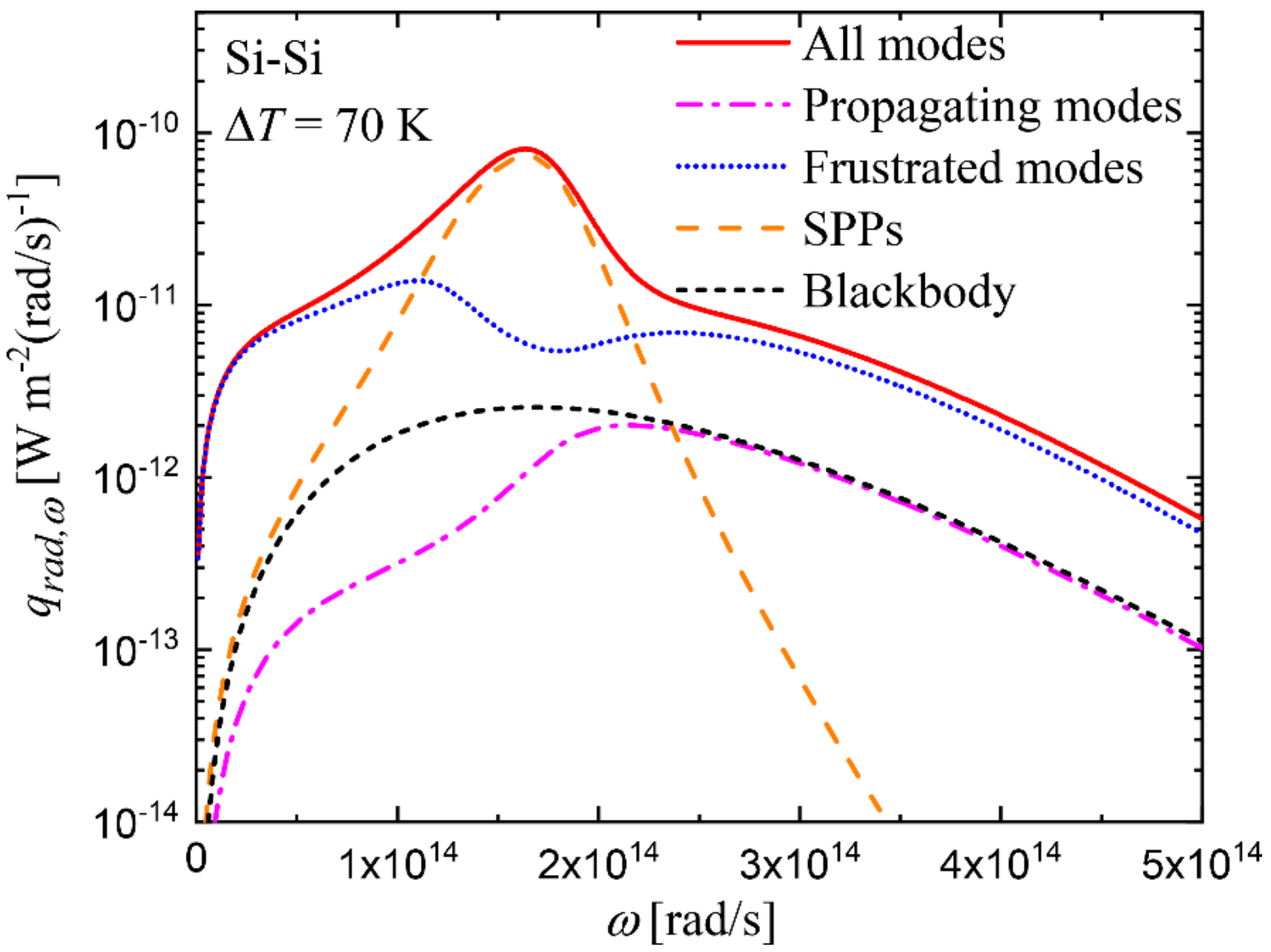

(b)

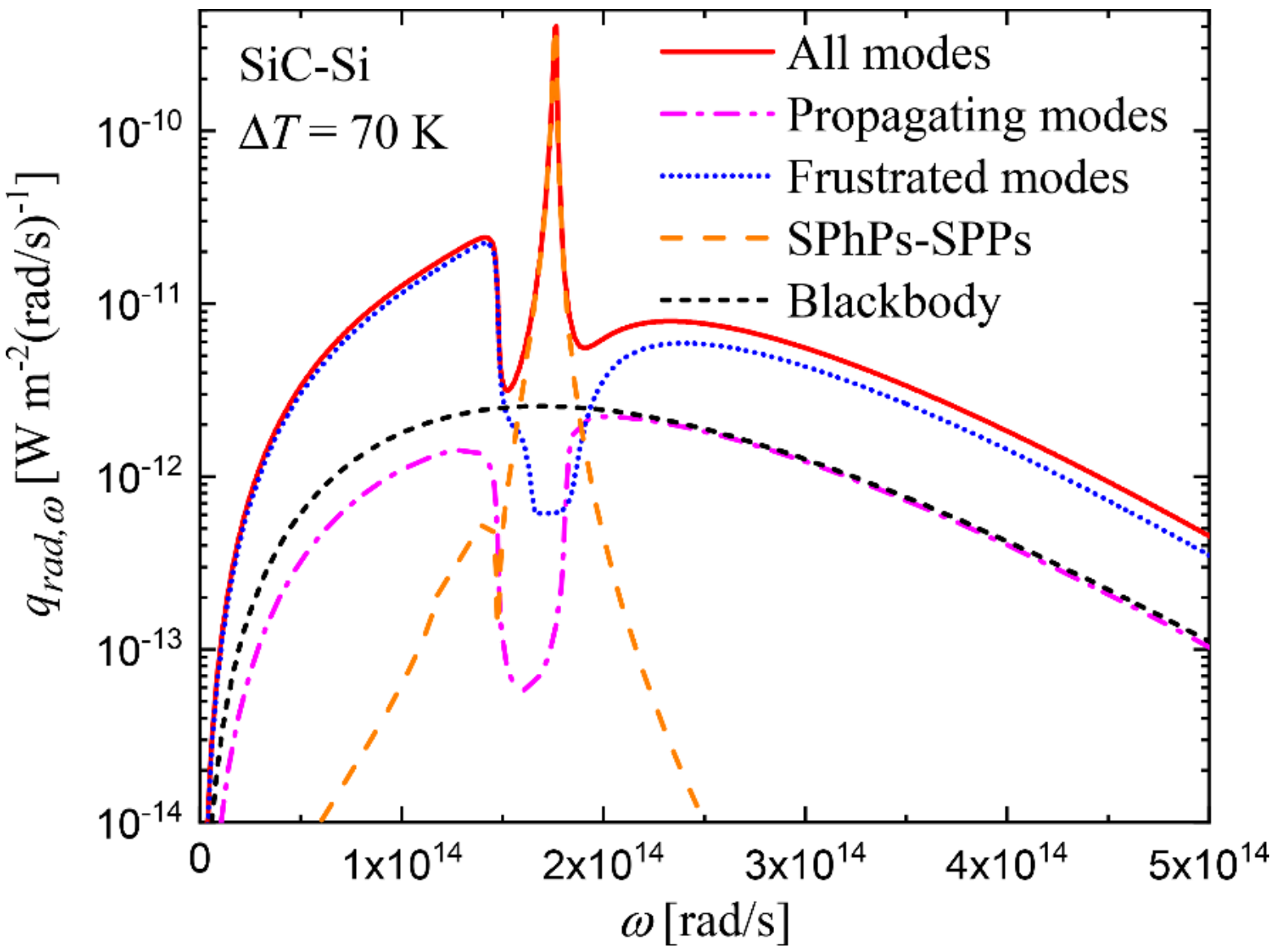

**Figure 4.** Radiative heat flux, $q_{rad,\omega}$, per unit angular frequency, $\omega$, for a vacuum gap spacing of 150 nm and a temperature difference of 70 K ($T_e$ = 370 K, $T_r$ = 300 K): (1) Si-Si sample. (b) SiC-Si sample. Separate contributions to the spectral radiative flux from propagating, frustrated and surface polariton modes are provided. The radiative flux between two blackbodies is also plotted for reference.

## Supporting Information

# Near-field radiative heat transfer between dissimilar materials mediated by coupled surface phonon- and plasmon-polaritons

Lei Tang[†], John DeSutter[‡], and Mathieu Francoeur[‡,*]

[†]Department of Mechanical Engineering, University of California, Berkeley, Berkeley, CA 94720, USA

[‡]Radiative Energy Transfer Lab, Department of Mechanical Engineering, University of Utah, Salt Lake City, UT 84112, USA

[*]Corresponding author: mfrancoeur@mech.utah.edu

## 1. Sample fabrication

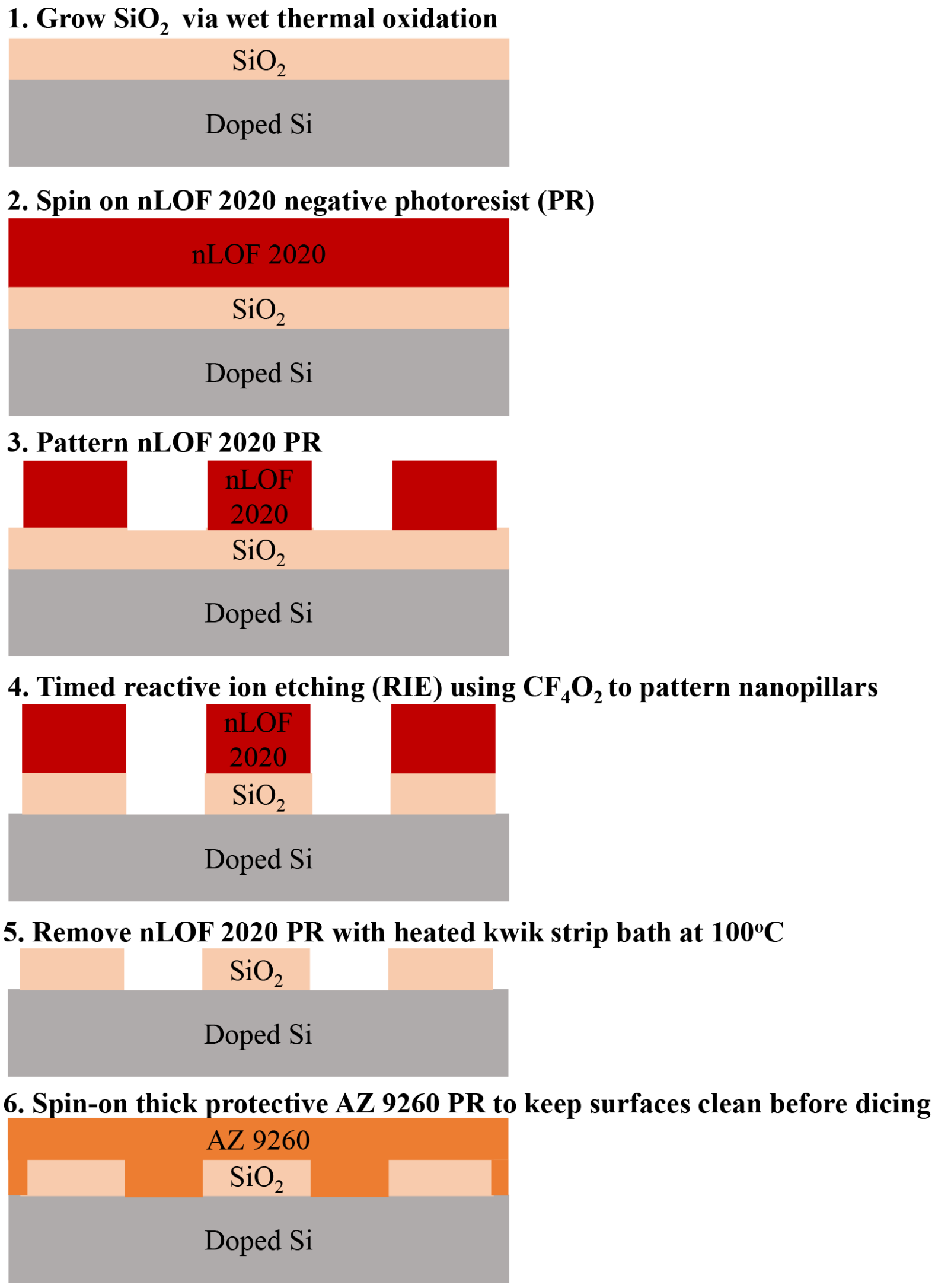

**Figure S1.** Main fabrication steps of the samples used for measuring near-field radiative heat transfer.

The key steps for manufacturing the $SiO_2$ nanopillars separating the emitter and receiver are summarized in Figure S1 and are described below:

1. A $SiO_2$ film is grown on a doped Si substrate (diameter of 100 mm) using a wet thermal oxidation in a furnace at 950ºC for 51 minutes.

2. A layer of nLOF 2020 negative photoresist (PR) is spin coated onto the $SiO_2$ film with a spin speed of 2000 RPM (acceleration of 500 RPM/s) for a duration of 60 seconds.

3. The nLOF 2020 is patterned by first exposing it to UV radiation shadowed by a photomask. The nLOF 2020 is then developed.

4. The $SiO_2$ nanopillars are patterned using a $CF_4O_2$ reactive ion etch (RIE).

5. The masking nLOF 2020 layer is removed using a Kwik strip bath at 100ºC followed by rinsing with acetone, isopropanol (IPA), and deionized (DI) water.

6. To avoid debris on the surface when dicing, a thick protective AZ 9260 PR layer (~ 10 to 15 μm) is deposited onto the wafer, and dicing tape is then adhered onto the protective layer. Subsequently, the wafer is diced into $5 \times 5$ mm$^2$ substrates using a Disco Dad641 dicing saw. Lastly, the diced surfaces are sonicated in acetone for ~ 5 min to remove the AZ 9260 PR layer and tape.

$SiO_2$ film thicknesses of 150 nm were measured via interferometry prior to RIE. As profilometry measurements demonstrate ~ 150-nm-tall nanopillars after etching (see SI Section 2), it is assumed that the gap spacing is maintained via nanopillars made exclusively of $SiO_2$.

The samples were carefully cleaned prior to performing near-field radiative heat transfer (NFRHT) experiments. The doped Si and 6*H*-SiC surfaces were first sonicated in acetone and IPA for 5 minutes and subsequently dried using a nitrogen blow gun. This was followed by careful inspection in a microscope (Olympus MX51). If a large amount of contaminants was detected, the surfaces were cleaned using solutions of piranha ($H_2O_2$:$H_2SO_4$ = 1:3) and standard clean 1 ($H_2O$:$H_2O_2$:$NH_4OH$ = 5:1:1 at 80°C) followed by spraying with acetone, IPA, and DI water. Only the spray cleaning procedure was utilized if a few particles were on the sample surfaces. Particles near edges of the surfaces can only be removed by using cleanroom wipes. Therefore, surface cleaning typically required iterations involving the aforementioned solutions/spray/wipe method and surface contaminant detection in the microscope. Once no visible particles were detected in the microscope, the receiver surface was placed in the vacuum chamber and aligned manually with the emitter. Specifically, the emitter was deposited onto the $SiO_2$ nanopillars without applying any external pressure and its position was slightly adjusted until it aligned well with the receiver.

Note that not all surfaces can be pristine, even when using the extensive cleaning procedure described above. Surfaces with even a single, small particle observed under the microscope typically resulted in invalid experimental results. Conversely, when NFRHT measurements were performed using clean surfaces (i.e., no visible particles were detected), a good agreement between experimental data and theoretical predictions were typically obtained.

**2. Gap spacing estimation**

The gap spacing between the emitter and receiver is determined based on the measured nanopillar heights and simulations for estimating the nanopillar and surface deflections.

The nanopillar heights are measured using a Tencor P-20H profilometer. Since profilometry measurements demonstrate that the nanopillar heights for both the SiC-Si and Si-Si samples are ~ 150 nm, profilometry data for only one nanopillar of the SiC-Si sample are shown in Figure S2. Small variations of the nanopillar heights are observed, and the minimum and maximum heights are 132 and 172 nm for both the SiC-Si and Si-Si samples. These measurements were performed before and after NFRHT experiments to ensure that the nanopillars were not damaged or plastically deformed during the experiments.

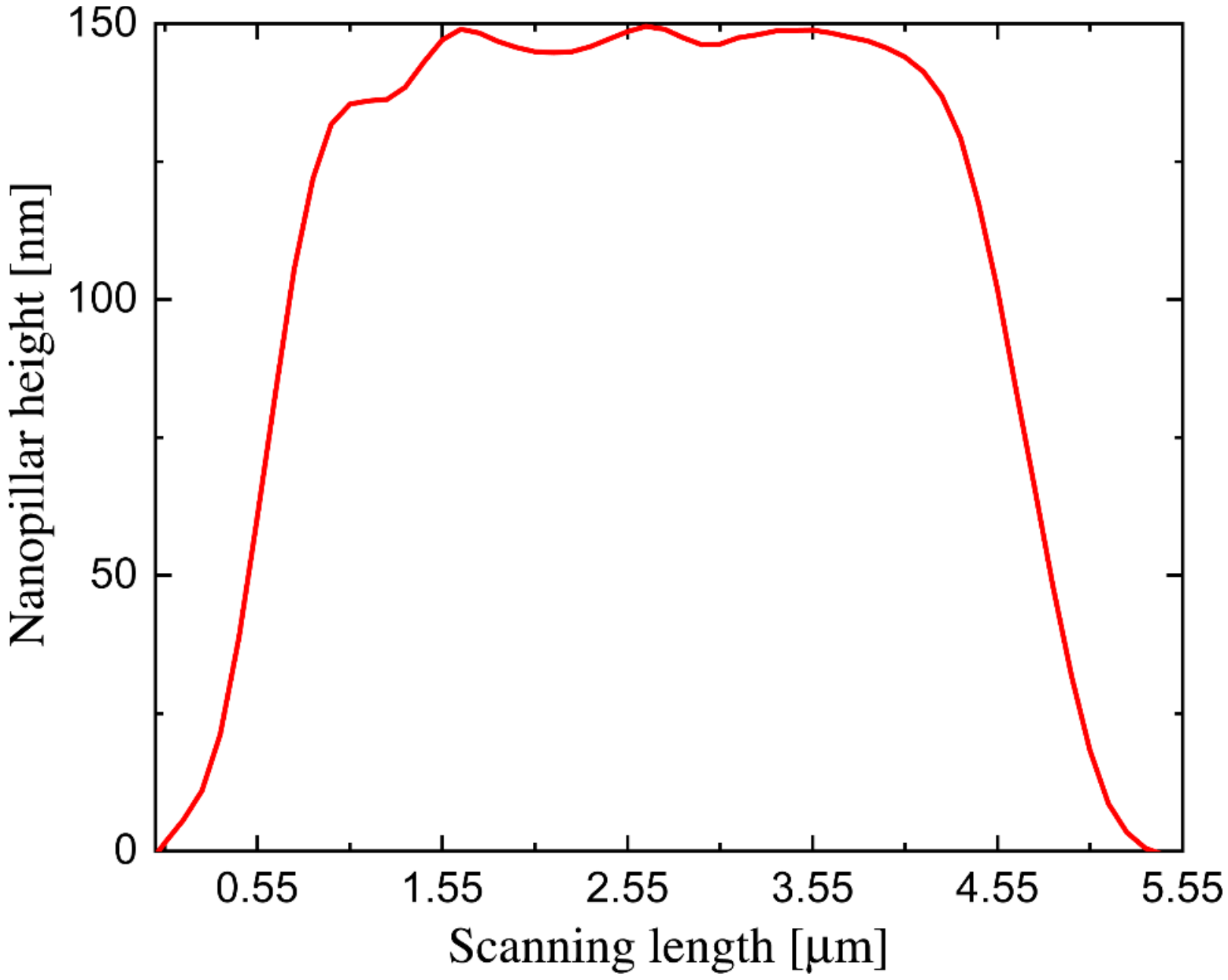

**Figure S2.** Measured nanopillar height as a function of the scanning length for the SiC-Si sample. The measured heights are similar for all nanopillars in both the SiC-Si and Si-Si samples. In addition, the measurements are similar before and after NFRHT experiments.

A uniform force was applied on the thermoelectric (TE) heater via a calibrated 10-g-mass in order to minimize the thermal contact resistances. The applied force can cause small elastic deflections of the nanopillars and substrates. Therefore, simulations for predicting such deflections are required. Only the analysis for the SiC-Si sample is provided hereafter since the same methodology was used for the Si-Si sample.

In order to determine the minimum gap spacing between the SiC and Si substrates, the deflection of the shortest possible nanopillar is first determined. Under a force exerted by a 10-g-mass, a maximum deflection of ~ 6 nm is predicted via COMSOL simulations for a 3-µm-diameter, 132-nm-tall $SiO_2$ nanopillar (see Figure S3). Therefore, the minimum gap spacing is obtained by subtracting a nanopillar deflection of 6 nm from the minimum nanopillar profilometry measurement of 132 nm.

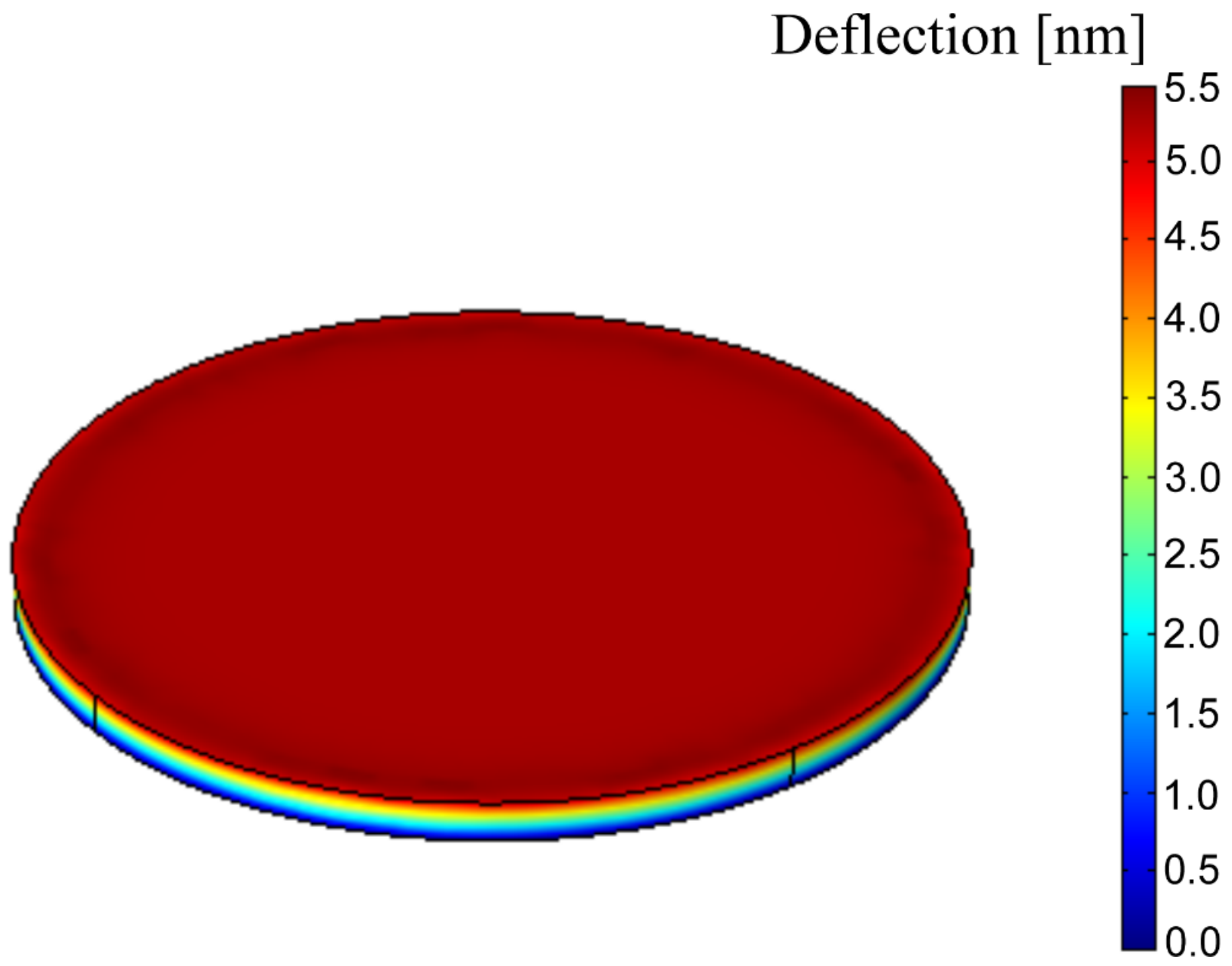

**Figure S3.** Deflection of a 3-µm-diameter, 132-nm-tall $SiO_2$ nanopillar under the force exerted by a 10-g-mass. Simulations are performed via COMSOL.

Next, the deflection of the substrates is considered. The substrate bows were first measured using a Tencor P-20H profilometer to scan across the entire surface of the substrates along two directions (0º and 90º scanning angles). The data for the largest bow between the 0º and 90º scanning angles were averaged by using parabolic fitting. For instance, the largest bow of the SiC substrate is along the 0º scanning angle. As such, the SiC bow was determined by performing a parabolic fitting based on the data at 0º (see Figure S4). From this fitted curve, a maximum height $H_1$ of ~ 50 nm is estimated to occur in the center of the SiC substrate, while the height $H_2$, measured from the substrate edge to the nanopillar location, is ~ 32 nm. The Si receiver has $H_1$ and $H_2$ values of 22 nm and 14 nm, respectively.

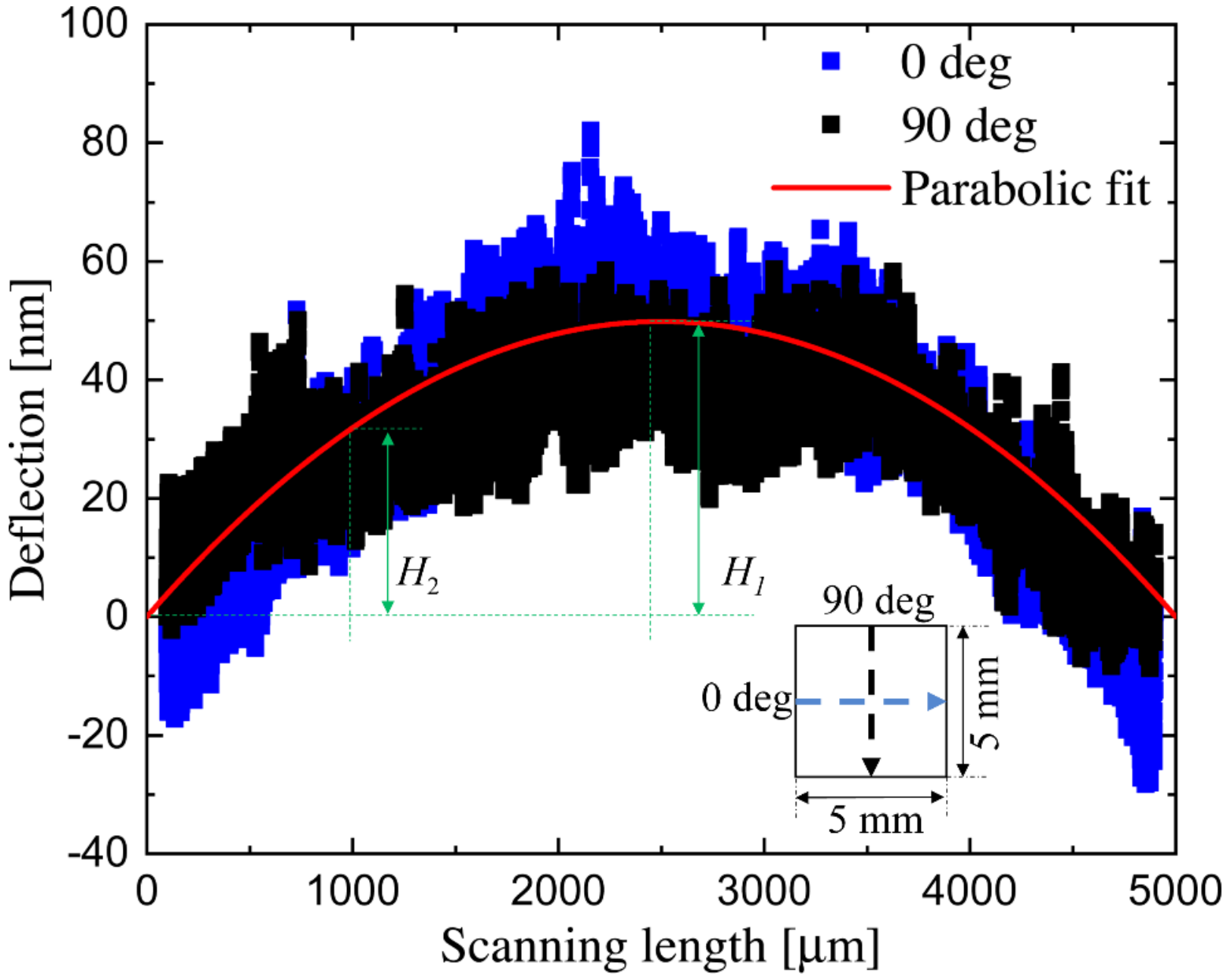

**Figure S4.** Measured bow of the SiC substrate. The bow is averaged by performing a parabolic fitting of the largest measurement.

Note that all bow measurements have been performed when the polished side of the substrates is facing up. Since the bow of both the SiC and Si substrates have a convex shape in this orientation, it is assumed that only three nanopillars, with one in the center and two on the left side of the Si surface, are in contact with the SiC emitter when no force is applied. Note that the polished side of the emitter is facing down after the sample is aligned. In addition to minimizing the thermal contact resistances, the force applied onto the TE heater ensure that the other two nanopillars on the right side of the Si surface are in contact with the emitter surface. COMSOL simulations were utilized for predicting the deflection of the emitter substrate by fixing three points corresponding to the locations where the nanopillars are in contact with the SiC substrate when no force is applied. Here, the nanopillar diameter is neglected since it is extremely small compared to the size of the substrate. For the two other nanopillars to be in contact with the emitter, the deflection of the SiC substrate at the locations where the nanopillars are not in contact must be ~ 52 nm, which is estimated using $2[(H_1 - H_2)_{SiC}+(H_1 - H_2)_{Si}]$. As shown in Figure S5, the deflection of the emitter at the location where the nanopillars are not in contact is ~ 40 nm, which is slightly smaller than 52 nm. A smaller deflection is obtained because a 315-μm-thick SiC substrate is assumed in the simulations. As provided by manufacturer (MTI, SC6HZ050503-033s1), the thickness of SiC varies from ~ 300 to 330 μm. It is therefore possible that the SiC substrate is slightly thinner than 315 μm, which would cause a deflection larger than 40 nm. For example, when the SiC substrate has a thickness of 300 μm, the deflection of the emitter at the location where the nanopillars are not in contact is ~ 51 nm, which is close to the desired value. Therefore, since the difference of 1 nm to 12 nm is very small, it is safe to conclude that all five nanopillars manufactured onto Si are in contact with the emitter under the force exerted by a 10-g-mass, which indicate that the SiC surface is deflected by 26 nm compared to the original bow during the experiment. Note that there is no further deflection of the sample when all nanopillars are in contact with the emitter as they are rigid. This was verified via COMSOL simulations. The largest possible gap spacing occurs near the edges of the samples, which is the height of the tallest possible nanopillar plus the sum of the bow of the Si and SiC substrates after the deflection. Specifically, an additional 46-nm-gap combined with the largest nanopillar height measurement (i.e., 172 nm) determines the largest possible gap spacing. As such, a gap spacing ranging from 126 nm to 218 nm (i.e., $d = 150^{+68}_{-24}$ nm) is determined for the SiC-Si sample and is used for the theoretical predictions. For the Si-Si sample, the gap spacing varies from 126 nm to 206 nm (i.e., $d = 150^{+56}_{-24}$ nm).

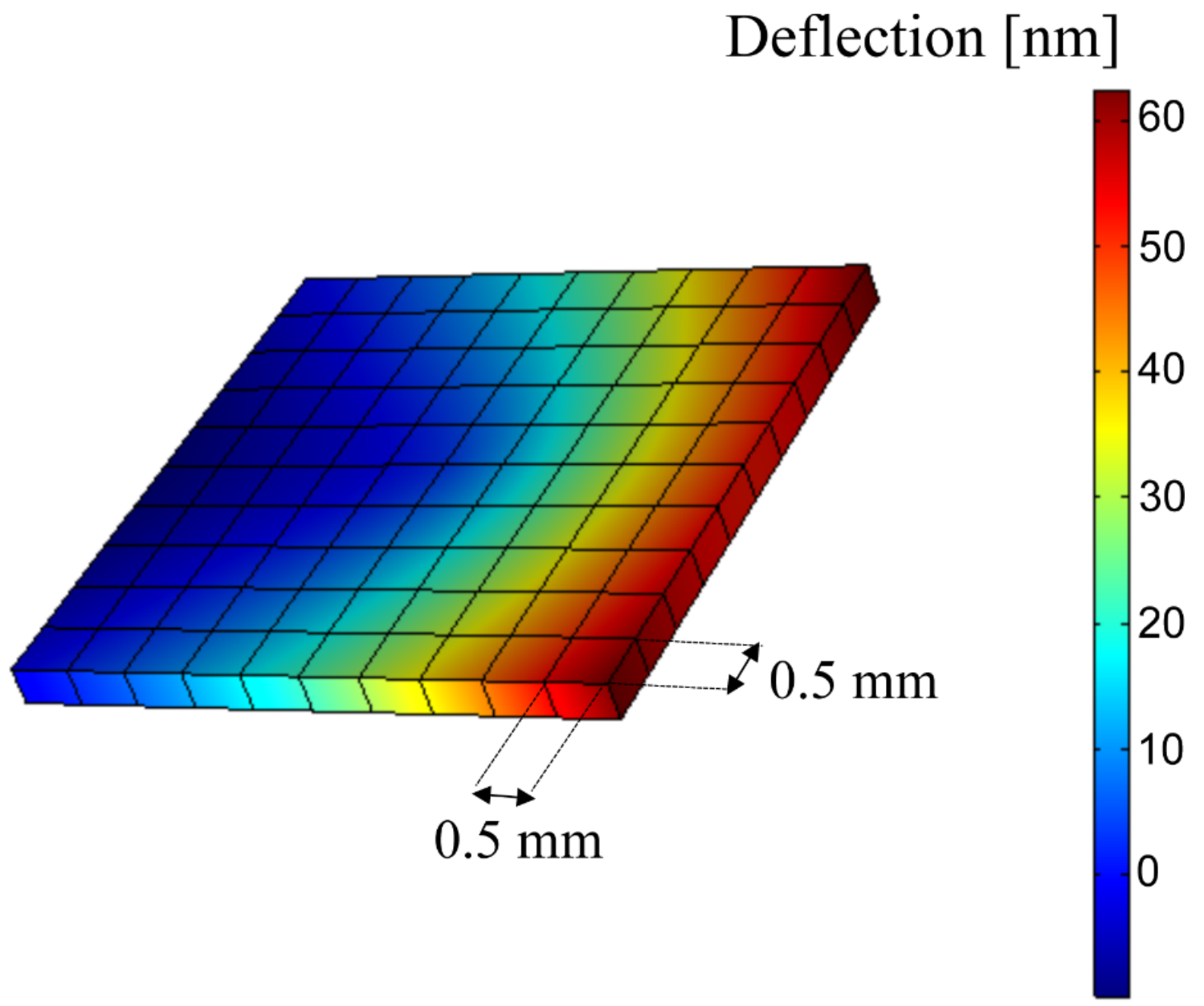

**Figure S5.** Deflection of the SiC emitter substrate under the force exerted by a 10-g-mass. The $SiO_2$ nanopillars are assumed to be points on the substrate. The nanopillars in the center and on the left side of the substrate are fixed (i.e., no deflection) when simulating the deflection of the emitter. Simulations are performed via COMSOL.

The simulated nanopillar deflection and measured bow of the SiC-Si and Si-Si samples are summarized in Table S.1.

**Table S1.** Details of the SiC-Si and Si-Si samples used for NFRHT measurements.

| Applied mass [g] | Vacuum gap spacing *d* [nm] | Sample | Bow | | Nanopillar location *S* [mm] | Number of nanopillars | Nanopillar diameter *D* [μm] | Nanopillar deflection [nm] |
|---|---|---|---|---|---|---|---|---|
| | | | $H_1$ [nm] | $H_2$ [nm] | | | | |
| 10 | 150 | 6*H*-SiC (emitter) | 50 | 32 | - | - | - | - |
| | | Highly doped Si (receiver) | 22 | 14 | 1 | 5 | 3 | 6 |
| 10 | 150 | Highly doped Si (emitter) | 22 | 19 | - | - | - | - |
| | | Highly doped Si (receiver) | 18 | 15 | 1.5 | 5 | 3 | 6 |

## 3. Uncertainty analysis

The colored bands of theoretical predictions for the measured heat rate (see Figure S6) are calculated based on the uncertainties introduced by the Si doping concentration, the sample bow, and the nanopillar height measurements. Si doping concentrations ranging from 4.3 to $4.9 \times 10^{19}$ $cm^{-3}$ were measured using a four-point-probe. The uncertainty of the nanopillar heights can affect the predictions of conduction heat transfer. Nanopillar heights were obtained via profilometry measurements. In addition, the nanopillar diameters were extracted based on SEM and Keyence microscope images. The nanopillar diameters were found to be consistent with the expected value of 3 μm, such that the effect of nanopillar diameter on conduction was neglected since it has no noticeable impact on the theoretical predictions. The calculated heat rate also accounts for the uncertainties due to the discrepancies of the nanopillar heights and sample bow, since the nanopillar height variations combined with the structural analysis were used to determine the gap spacings as discussed in SI Section 2. Note that the experimental data of radiative flux are obtained by subtracting theoretical conduction from the heat rate measured with the heat flux meter (HFM), such that the uncertainty in the amount of conduction is included in the experimental data shown in Figure 2.

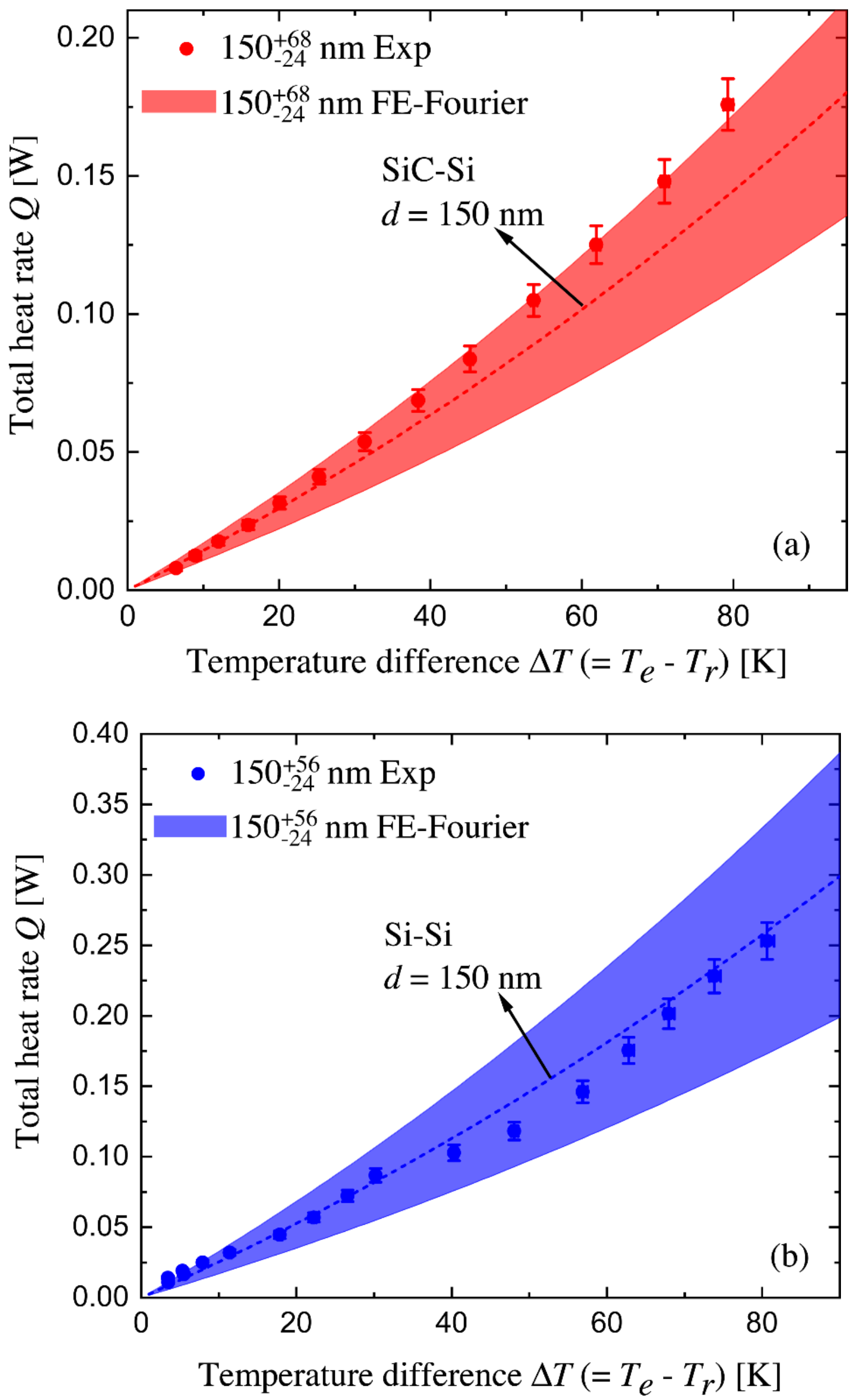

**Figure S6.** Total heat transfer rate, $Q$, as a function of the temperature difference, $\Delta T = T_e - T_r$: (a) SiC-Si sample. (b) Si-Si sample. Experimental results (Exp) are shown by symbols, and are compared against theoretical predictions based on fluctuational electrodynamics (FE) and Fourier's law. The colored band of theoretical predictions are calculated based on the nanopillar height, sample bow and Si doping level. In all experiments, the temperature of the receiver, $T_r$, is kept at ~ 300 K.

The upper (lower) theoretical curve of heat rate and flux is calculated based on the smallest (largest) possible gap spacing, the Si doping concentration providing the largest (smallest) heat transfer, and the largest (smallest) amount of conduction calculated with the smallest (largest) nanopillar height. For example, the upper theoretical curve of heat rate for the SiC-Si sample was first determined by calculating NFRHT using the smallest possible gap spacing (126 nm) and the largest Si doping concentration ($4.9 \times 10^{19}$ $cm^{-3}$). Then, the maximum conduction was obtained by using Fourier's law based on the smallest possible nanopillar height (126 nm). Finally, the upper theoretical curve was generated by summing the calculated NFRHT and conduction.

The uncertainties of the experimental data come from the temperature and heat rate measurements. The uncertainty of the temperature measurements firstly stems from the ohmmeter used to probe the resistance of the thermistors. The uncertainty induced from such device is small (less than ± 0.05 K). The details of this uncertainty can be found in Refs. 1 and 2. Additionally, the thermistors have an accuracy of ± 0.1 K according to the manufacturer. The more pronounced uncertainty associated with temperature measurements comes from the uncertainty introduced by the thermal resistance of the grease. The measured thermal resistance of the grease can be found in Refs. 2 and 3. By accounting for all of these factors, the maximum and minimum uncertainties on the temperature differences for the SiC-Si sample are respectively $79.3^{+0.7}_{-0.6}$ K and $6.4^{+0.2}_{-0.2}$ K, while the maximum and minimum uncertainties for the Si-Si sample are $80.6^{+0.9}_{-0.7}$ K and $3.5^{+0.2}_{-0.2}$ K, respectively.

The uncertainty of the heat rate measurement stems from the HFM. The HFM has an error of ± 5%, as provided by the manufacturer. Since each experimental point is the average of a set of data, additional errors are introduced by the distribution of each dataset. These distribution errors associated with each dataset are accounted for by taking two standard deviations of the mean. Note that each dataset was obtained from the HFM by recording values every second for at least two minutes once the heat transferred between the two surfaces reached steady state. By combining these uncertainties, the largest and smallest uncertainties in the measured heat rate for the SiC-Si sample are respectively $0.176^{+0.009}_{-0.009}$ W for a temperature difference of ~ 79.3 K and $0.008^{+0.001}_{-0.001}$ W for a temperature difference of ~ 6.4 K. For the Si-Si sample, the largest and smallest uncertainties of heat rate are $0.253^{+0.013}_{-0.013}$ W for a temperature difference of ~ 80.6 K and $0.011^{+0.001}_{-0.001}$ W for a temperature difference of ~ 3.5 K, respectively. These uncertainties are included as error bars in the experimental data shown in Figure 2 (radiative flux) and Figure S6 (heat rate).